\documentclass[
    ,final            
  ,nomathfonts                 
  ]
  {aipproc}

\layoutstyle{6x9}

\newcommand{\be}{\begin{equation}}
\newcommand{\ee}{\end{equation}}
\newcommand{\bea}{\begin{eqnarray}}
\newcommand{\eea}{\end{eqnarray}}

\newcommand{\aeq}{&=&}

\newcommand{\itDelta}{{\it \Delta}}

\newcommand{\itLambda}{{\Lambda}}

\newcommand{\itPi}{{\Pi}}

\newcommand{\bra}{\langle}
\newcommand{\ket}{\rangle}
\newcommand{\dbra}{\bra \! \bra}
\newcommand{\dket}{\ket \! \ket}

\newcommand{\me}{\mbox{e}}

\newcommand{\bq}{{\bar q}}

\newcommand{\rRe}{{\rm Re}}

\newcommand{\rS}{{\rm S}}

\begin{document}

\title{Multifractal Analysis of Various PDF in Turbulence based on
Generalized Statistics:
A Way to Tangles in Superfluid He}

\author{Toshihico Arimitsu}{
  address={Institute of Physics, University of Tsukuba,
Ibaraki 305-8571, Japan}
,altaddress={E-mail: arimitsu@cm.ph.tsukuba.ac.jp}
}

\author{Naoko Arimitsu}{
  address={Graduate School of EIS, Yokohama Nat'l.~University, 
Yokohama 240-8501, Japan}
}


\begin{abstract}
By means of the multifractal analysis (MFA), the expressions of
the probability density functions (PDFs) are unified in a compact analytical
formula which is valid for various quantities in turbulence. 
It is shown that the formula
can explain precisely the experimentally observed PDFs both on log and
linear scales.
The PDF consists of two parts, i.e., the {\it tail} part and the {\it center} part.
The structure of the tail part of the PDFs, determined mostly by
the intermittency exponent, represents the intermittent large deviations 
that is a manifestation of the multifractal distribution of singularities 
in physical space due to the scale invariance of the Navier-Stokes equation 
for large Reynolds number. 
On the other hand, the structure of the center part represents small deviations 
violating the scale invariance due to thermal fluctuations and/or observation error.


\end{abstract}

\maketitle


\section{Introduction}
\label{intro}

In this paper, we derive the unified formula for various 
probability density functions (PDFs) 
in fully developed turbulence by means of 
the {\it multifractal analysis} (MFA)
\cite{AA,AA1,AA2,AA3,AA4,AA5,AA6,AA7,AA8,AA9,AA10,AA11}, 
and analyze the PDFs observed in two experiments, i.e., 
the PDFs of velocity fluctuations,
of velocity derivatives and of fluid particle accelerations 
at $R_\lambda = 380$ that was extracted by Gotoh et al.\ from
the DNS of the size 1024$^3$ \cite{Gotoh02}, and 
the PDF of fluid particle accelerations at $R_\lambda = 690$ obtained in 
the Lagrangian measurement of particle accelerations 
that was realized by Bodenschatz and co-workers \cite{EB01a,EB01b,EB02comment} 
by raising dramatically the spatial and temporal measurement resolutions
with the help of the silicon strip detectors.
The MFA of turbulence 
is a unified self-consistent approach for the systems with large deviations,
which has been constructed based on 
the Tsallis-type distribution function \cite{Tsallis88,Tsallis99}
that provides an extremum of the {\it extensive} R\'{e}ny \cite{Renyi} 
or the {\it non-extensive} Tsallis entropy \cite{Tsallis88,Tsallis99,Havrda-Charvat}
under appropriate constraints.
The analysis rests
on the scale invariance of the Navier-Stokes equation for high Reynolds number, 
and on the assumptions that the singularities due to the invariance 
distribute themselves multifractally in physical space.
The MFA is a generalization of 
the log-normal model \cite{Oboukhov62,K62,Yaglom}. It has been shown \cite{AA4} that
the MFA derives the log-normal model 
when one starts with the Boltzmann-Gibbs entropy.

For high Reynolds number $\rRe \gg 1$, or for the situation where 
effects of the kinematic viscosity $\nu$ can be neglected compared with
those of the turbulent viscosity, the Navier-Stokes equation,
$
\partial {\mathbf u}/\partial t
+ ( {\mathbf u}\cdot {\mathbf \nabla} ) {\mathbf u} 
= - {\mathbf \nabla} \left(p/\rho \right)
+ \nu \nabla^2 {\mathbf u}
\label{N-S eq}
$,
of an incompressible fluid is invariant under 
the scale transformation~\cite{Moiseev76,Frisch-Parisi83,Meneveau87b}
$
{\mathbf r} \rightarrow \lambda {\mathbf r}
$, 
$
{\mathbf u} \rightarrow \lambda^{\alpha/3} {\mathbf u}
$, 
$
t \rightarrow \lambda^{1- \alpha/3} t
$ and
$
\left(p/\rho\right) \rightarrow \lambda^{2\alpha/3} \left(p/\rho\right)
$
where the exponent $\alpha$ is an arbitrary real quantity.
The quantities $\rho$ and $p$ represent, respectively, mass density and pressure.
The Reynolds number $\rRe$ of the system is given by 
$
{\rm Re} = \delta u_{\rm in} \ell_{\rm in}/\nu = ( \ell_{\rm in}/\eta )^{4/3}
$
with the Kolmogorov scale 
$
\eta = ( \nu^3/\epsilon )^{1/4}
$~\cite{K41}
where $\epsilon$ is the energy input rate at the input scale 
$\ell_{\rm in}$.
Here, we introduced 
$\delta u_{\rm in} = \vert u(\bullet + \ell_{\rm in}) - u(\bullet) \vert
$ 
with the definition of the velocity fluctuation (difference)
$
\delta u_n = \vert u(\bullet + \ell_n) - u(\bullet) \vert
$
where $u$ is a component of velocity field $\mathbf{u}$, and
$\ell_n$ is a distance between two points.
The {\it pressure} (divided by the mass density) difference 
$
\delta p_n = \vert p/\rho(\bullet + \ell_n) - p/\rho(\bullet) \vert
$
between two points separated by the distance $\ell_n$ is another
important observable quantity.
We are measuring distance by the discrete units 
$
\ell_n = \delta_n \ell_0
\label{r-n}
$
with $\delta_n = 2^{-n}$ $(n=0,1,2,\cdots)$.
The non-negative integer $n$ can be interpreted as 
the {\it multifractal depth}.
However, we will treat it as positive real number in the analysis of experiments.
The multifractal depth $n$ is related to 
the number of steps within the energy cascade model.\footnote{
The definition of the number of steps $\bar{n}$ 
within the energy cascade model is given by
$
\bar{n}= -\log_2 (r /\ell_{\rm in})
$
for the eddies whose diameter is equal to $r$.
By putting $r = \ell_n$, this gives us 
the relation between $\bar{n}$ and $n$ in the form
\be
\bar{n} = n - \log_2 (\ell_0/\ell_{\rm in}).
\label{n-nbar}
\ee
}
At each step of the cascade, say at the $n$th step, eddies break up into 
two pieces producing an energy cascade with the energy-transfer rate
$\epsilon_n$ that represents the rate of transfer of energy per unit mass 
from eddies of size $\ell_n$ to those of size $\ell_{n+1}$.

\section{Singularities and Scaling exponents}

Let us consider the quantity 
$
\delta x_n = \vert x(\bullet + \ell_n) - x(\bullet) \vert
$
having the scaling property
$
\vert x_n \vert \equiv \vert \delta x_n / \delta x_0 \vert
= \delta_n^{\phi \alpha/3}
$.
Its spatial derivative is defined by
$
\vert x' \vert = \lim_{\ell_n \rightarrow 0} \delta x_n /\ell_n 
\propto \lim_{n \rightarrow \infty} \ell_n^{\phi \alpha /3 -1}
$
which becomes singular for 
$
\alpha < 3/\phi
$.
The values of exponent $\alpha$ specify the degree of singularity.
We see that the scale invariance provides us with
$
\delta u_n /\delta u_0 = \delta_n^{\alpha/3}
$
and
$
\delta p_n / \delta p_0 = (\ell_n / \ell_0)^{2\alpha/3}
\label{p-alpha}
$
giving, respectively,
$
\phi = 1
$
for the velocity fluctuation and
$
\phi = 2
$
for the pressure fluctuation.
The velocity derivative and 
the fluid particle acceleration may be estimated, respectively, by 
$
\vert u^\prime \vert = \lim_{n \rightarrow \infty} u^\prime_n
$
and by
$
\vert \mathbf{a} \vert = \lim_{n \rightarrow \infty} \mathrm{a}_n
$
where we introduced the $n$th velocity derivative
$
u'_n = \delta u_n / \ell_n
$
and the $n$th fluid particle acceleration 
$
\mathrm{a}_n = \delta p_n / \ell_n
$
corresponding to the characteristic length $\ell_n$.
Note that the acceleration $\mathbf{a}$ of a fluid particle is given by
the substantive time derivative of the velocity:
$
{\mathbf a} = \partial {\mathbf u}/\partial t
+ ( {\mathbf u}\cdot {\mathbf \nabla} ) {\mathbf u}
$.
We see that the velocity derivative and 
the fluid particle acceleration become singular for $\alpha < 3$ and $\alpha < 1.5$, 
respectively,
i.e.,
$
\vert u' \vert \propto \lim_{\ell_n \rightarrow 0} \ell_n^{(\alpha/3)-1}
\rightarrow \infty
$
and
$
\vert \mathbf{a} \vert \propto \lim_{\ell_n \rightarrow 0} \ell_n^{(2\alpha/3)-1}
\rightarrow \infty
$.
We also see that the energy dissipation rate becomes singular in the limit 
$n \rightarrow \infty$ for $\alpha < 1$,
i.e.,
$
\lim_{n \rightarrow \infty} \epsilon_n / \epsilon_0 
= \lim_{n \rightarrow \infty} ( \ell_n/\ell_0 )^{\alpha -1}
\rightarrow \infty
$
giving $\phi = 3$.

\begin{figure}
  \includegraphics[height=.25\textheight]{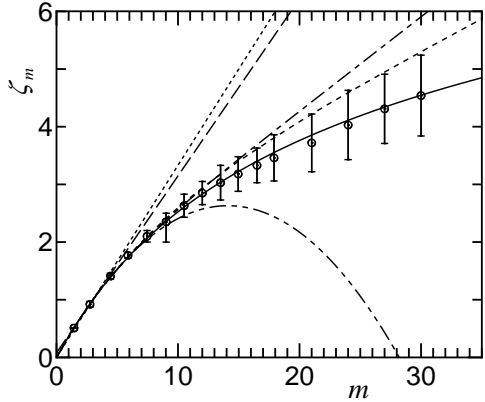}
  \caption{Comparison of the present scaling exponents $\zeta_m$ for 
  $\mu=0.238$ (solid curve) with the experimental results plotted by 
  circles at $R_{\lambda}=110$ ($\rRe = 32000$)~\cite{Meneveau91}, and
  with other theories with the same value of $\mu$. K41 is given by the dotted line, 
  $\beta$-model by the dashed line, p-model by the dotted dashed line,
  log-Poisson model by the short dashed curve, and 
  log-normal by the two dotted dashed curve.}
\label{zeta-m Benzi95}
\end{figure}

The MFA rests on the multifractal distribution of 
singularities that is a manifestation of the scale invariance of the
Navier-Stokes equation for large Re as mentioned above.
The probability 
$
P^{(n)}(\alpha) d\alpha
$
to find, at a point in physical space, a singularity labeled by an exponent 
in the range 
$
\alpha \sim \alpha + d \alpha
$
is given by~\cite{AA1,AA2,AA3,AA4}
$
P^{(n)}(\alpha) = \left[ 1 - (\alpha - \alpha_0)^2 \big/ (\itDelta \alpha )^2 
\right]^{n/(1-q)}/Z_{\alpha}^{(n)} 
\label{Tsallis prob density}
$
with an appropriate partition function
$
Z_{\alpha}^{(n)}
$
and
$
(\itDelta \alpha)^2 = 2X \big/ [(1-q) \ln 2 ]
$.
This is consistent with the relation \cite{Meneveau87b,AA4}
$
P^{(n)}(\alpha) \propto \delta_n^{1-f(\alpha)}
$
that reveals how densely each singularity, labeled by $\alpha$, 
fills physical space.
In the present model, the multifractal spectrum $f(\alpha)$
is given by~\cite{AA1,AA2,AA3,AA4}
$
f(\alpha) = 1 + (1-q)^{-1} \log_2 [ 1 - (\alpha - \alpha_0 )^2
/ (\Delta \alpha )^2 ]
\label{Tsallis f-alpha}
$.
The range of $\alpha$ is $\alpha_{\rm min} \leq \alpha \leq \alpha_{\rm max}$ with
$
\alpha_{\rm min} = \alpha_0 - \itDelta \alpha
$, 
$
\alpha_{\rm max} = \alpha_0 + \itDelta \alpha
$.
The distribution function $P^{(n)}(\alpha)$ is determined by taking an extremum of generalized 
entropies, i.e., the {\it extensive} R\'enyi entropy or the {\it non-extensive}
Tsallis entropy, under the condition that the information one has for the system
is only the value of the intermittency exponent.
In spite of the different characteristics of the entropies,
the distribution functions $P^{(n)}(\alpha)$ giving their extremum
have the common structure.\footnote{
Within the present formulation, the decision cannot be pronounced 
which of the entropies is underlying the system of turbulence.
}

The dependence of the parameters $\alpha_0$, $X$ and $q$ on 
the intermittency exponent $\mu$ is determined, 
self-consistently, with the help of the three independent equations, i.e.,
the energy conservation:
$
\left\bra \epsilon_n \right\ket = \epsilon
\label{cons of energy}
$,
the definition of the intermittency exponent $\mu$:
$
\bra \epsilon_n^2 \ket 
= \epsilon^2 \delta_n^{-\mu}
\label{def of mu}
$,
and the scaling relation~\footnote{
The scaling relation is a generalization of the one derived first in
\cite{Costa,Lyra98} to the case where the multifractal spectrum
has negative values.
}:
$
1/(1-q) = 1/\alpha_- - 1/\alpha_+
\label{scaling relation}
$
with $\alpha_\pm$ satisfying $f(\alpha_\pm) =0$. 
The average $\bra \cdots \ket$ is taken with $P^{(n)}(\alpha)$.

The scaling exponents $\zeta_m$ of the $m$th order velocity fluctuations,
defined by 
$
\bra \vert u_n \vert^m \ket = \bra \delta_n^{m\alpha/3} \ket \propto \delta_n^{\zeta_m}
$,
are given in the analytical form \cite{AA1,AA2,AA3,AA4}
\be
\zeta_m = \alpha_0 m/3 
- 2Xm^2/[9 (1+C_{m/3}^{1/2} )]
- [1-\log_2 (1+C_{m/3}^{1/2} ) ] /(1-q)
\label{zeta}
\ee
with
$
{C}_{\bq} = 1 + 2 \bq^2 (1-q) X \ln 2
\label{cal D}
$.
The formula (\ref{zeta}) is independent of $n$,
that is a manifestation of the scale invariance.

The derived scaling exponents (\ref{zeta}) are shown
in Fig.~\ref{zeta-m Benzi95} by the solid curve for the case $\mu = 0.238$,
and are compared with experimental data~\cite{Meneveau91} and 
with the curves given by other theories, i.e., 
K41~\cite{K41}, log-normal~\cite{Oboukhov62,K62,Yaglom}, 
$\beta$-model~\cite{Frisch78}, p-model~\cite{Meneveau87a,Meneveau87b}
and log-Poisson~\cite{She94,She95}.

\section{Various probability density functions}

It has been shown that the probability 
$\itPi^{(n)}_{\phi,\rS}(x_n) dx_n$ to find a physical quantity
$
x_n
$
in the range $x_n \sim x_n+dx_n$ is given in the form
\be
\Pi_\phi^{(n)}(x_n) dx_n = \Pi^{(n)}_{\phi,\rS}(x_n) dx_n 
+ \Delta \Pi_\phi^{(n)}(x_n) dx_n
\label{def of Pi phi}
\label{def of Lambda}
\ee
with the normalization
$
\int_{-\infty}^{\infty} dx_n  \Pi_\phi^{(n)}(x_n) =1
$.
The first term represents the contribution by the singular part of the quantity
$ x_n $ stemmed from the multifractal distribution of its singularities 
in physical space. This is given by
$
\Pi^{(n)}_{\phi,\rS}(\vert x_n \vert) dx_n \propto P^{(n)}(\alpha) d \alpha
\label{singular portion}
$
with the transformation of the variables,
$
\vert x_n \vert = \delta_n^{\phi \alpha/3}
$.
Whereas the second term $\Delta \Pi_\phi^{(n)}(x_n) dx_n$ represents 
the contribution from the dissipative term
in the Navier-Stokes equation, and/or the one 
from the errors in measurements.
The dissipative term has been discarded in the above investigation for 
the distribution of singularities since it violates 
the invariance under the scale transformation.
The contribution of the second term provides a correction to the first one.
Note that each term in (\ref{def of Lambda}) is a multiple of two 
probability functions, i.e., the one to
determine the portion of the contribution among the above mentioned two independent 
origins, and the other to find $x_n$ in the range $x_n \sim x_n+dx_n$.
Note also that the values of $x_n$ originated in the singularity
are rather large representing intermittent large deviations, and that those 
contributing to the correction terms are of the order of or smaller than
its standard deviation.

\begin{figure}
  \includegraphics[height=.35\textheight]{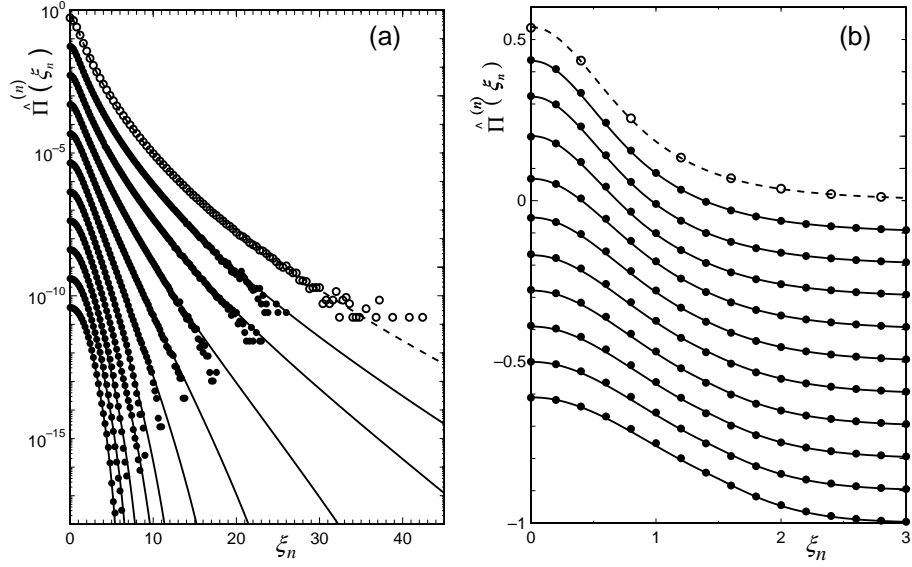}
  \caption{Analyses of the PDFs of {\it velocity fluctuations} (closed circles) and 
of {\it velocity derivatives} (open circles) measured in the DNS by Gotoh et al.\ 
at $R_\lambda = 380$ by the present theoretical PDFs $\hat{\itPi}^{(n)}(\xi_n)$ 
for {\it velocity fluctuations} (solid lines) and 
for {\it velocity derivatives} (dashed line) are plotted 
on (a) log and (b) linear scales.
The DNS data points are symmetrized by taking averages of 
the left and the right hand sides data.
The measuring distances, $r/\eta = \ell_n/\eta$, for the PDF of velocity fluctuations
are, from the second top to bottom: 2.38, 4.76, 9.52, 19.0, 38.1, 76.2, 152, 305, 
609, 1220.
For the theoretical PDFs of velocity fluctuations, 
$\mu = 0.240$ ($q=0.391$), from the second top to bottom: $(n,\ \bar{n},\ q') =$
(20.7,\ 14.6,\ 1.60), (19.2,\ 13.1,\ 1.60), (16.2,\ 10.1,\ 1.58), 
(13.6,\ 7.54,\ 1.50), (11.5,\ 5.44,\ 1.45), (9.80,\ 3.74,\ 1.40), 
(9.00,\ 2.94,\ 1.35), (7.90,\ 1.84,\ 1.30), (7.00,\ 0.94,\ 1.25), 
(6.10,\ 0.04,\ 1.20), 
$
\xi_n^* = 1.10 \sim 1.43
$
($\alpha^* = 1.07$), and
$
\xi_n^{\rm max} = 204 \sim 6.63
$.
For the theoretical PDF of velocity derivatives, 
$(n,\ \bar{n},\ q')=(22.4,\ 16.3,\ 1.55)$,
$
\xi_n^* = 1.06
$
($\alpha^* = 1.07$), and
$
\xi_n^{\rm max} = 302
$.
For better visibility, each PDF is shifted by $-1$ unit along the vertical axis.
}
\label{velocity fluctuations log}
\end{figure}

The $m$th moment of the variable $\vert x_n \vert$ is given by 
$
\dbra \vert x_n \vert^m \dket_{\phi} \equiv \int_{-\infty}^{\infty} dx_n  
\vert x_n \vert^m \itPi_\phi^{(n)}(x_n)
$$
= 2 \gamma^{(n)}_{\phi,m}
+ (1-2\gamma^{(n)}_{\phi,0} ) a_{\phi m}\ \delta_n^{\zeta_{\phi m}}
\label{structure func m}
$
where
$
2\gamma^{(n)}_{\phi,m} = \int_{-\infty}^{\infty} dx_n 
\vert x_n \vert^m \Delta \Pi_\phi^{(n)}(x_n),
$
and
$
a_{3\bq} = \{ 2 / [\sqrt{C_{\bq}} ( 1+ \sqrt{C_{\bq}} ) ] \}^{1/2}
$.

We now derive the PDF, $\hat{\itPi}_{\phi}^{(n)}(\xi_n)$, 
defined by the relation 
$
\hat{\Pi}_\phi^{(n)}(\xi_n) d\xi_n
= \Pi_\phi^{(n)}(x_n) d x_n
$
with the variable $\xi_n = x_n/\dbra \vert x_n \vert^2 \dket_{\phi}^{1/2}$ normalized by 
the standard deviation $\dbra x_n^2 \dket_{\phi}^{1/2}$. This PDF is to be compared with
the observed PDFs. The variable is related with $\alpha$ by
$
\vert \xi_n \vert = \bar{\xi}_n \delta_n^{\phi \alpha /3 -\zeta_{2\phi}/2}
$
with
$
\bar{\xi}_n = [2 \gamma_{\phi,2}^{(n)} \delta_n^{-\zeta_{2\phi}} 
+ (1-2\gamma_{\phi,0}^{(n)} ) a_{2\phi} ]^{-1/2}
$.
It is reasonable to imagine that the origin of intermittent rare events is 
attributed to the first singular term in (\ref{def of Lambda}), and that
the contribution from the second term is negligible. We then have for the tail part,
i.e., 
$\xi_n^* \leq \vert \xi_n \vert \leq \xi_n^{\rm max}$,
\bea
\hat{\Pi}_\phi^{(n)}(\xi_n) d \xi_n 
\aeq \Pi^{(n)}_{\phi,\rm S} (x_n) dx_n
\nonumber\\
\aeq \bar{\Pi}_\phi^{(n)} \frac{\bar{\xi}_n}{\vert \xi_n \vert}
\left[1 - \frac{1-q}{n}\ 
\frac{\left(3 \ln \vert \xi_n / \xi_{n,0} \vert\right)^2}{
2\phi^2 X \vert \ln \delta_n \vert} \right]^{n/(1-q)} d \xi_n
\label{PDF kappa large}
\eea
with
$
\bar{\Pi}_\phi^{(n)} = 3 (1-2\gamma^{(n)}_0)
/ (2\phi \bar{\xi}_n \sqrt{2\pi X \vert \ln \delta_n \vert} )
$,
$
\xi_{n,0} = \bar{\xi}_n \delta_n^{\phi \alpha_0 /3 -\zeta_{2\phi} /2}
$,
$
\xi_n^{\rm max} = \bar{\xi}_n \delta_n^{\phi \alpha_{\rm min}/3 -\zeta_{2\phi} /2}
$.
On the other hand, for the center part, 
the contribution to the PDF comes, mainly, from thermal fluctuations 
or measurement error. It may be described by the Tsallis distribution function
with respect to the variable $\xi_n$ itself, i.e.,
$
\vert \xi_n \vert \leq \xi_n^*
$,
\bea
\hat{\Pi}_\phi^{(n)}(\xi_n) d \xi_n \aeq
\left[ \hat{\Pi}^{(n)}_{\phi,\rS}(x_n)
+\Delta \hat{\Pi}_\phi^{(n)}(x_n) \right] d x_n
\nonumber\\
\aeq \bar{\Pi}_\phi^{(n)}
\left\{1-\frac{1-q'}{2} \left(1+\frac{3f'(\alpha^*)}{\phi}\right)
\left[ \left(\frac{\xi_n}{\xi_n^*}\right)^2 -1 \right] \right\}^{1/(1-q')} d \xi_n.
\label{PDF kappa small}
\eea
This specific form of the Tsallis distribution function is determined by the condition that 
the two PDFs (\ref{PDF kappa large}) and (\ref{PDF kappa small}) 
should have the same value and the same slope at
$\xi_n^*$ which is defined by
$
\xi_n^* = \bar{\xi}_n \delta_n^{\phi \alpha^* /3 -\zeta_{2\phi} /2}
$
with $\alpha^*$ being the smaller solution of 
$
\zeta_{2\phi}/2 -\phi \alpha/3 +1 -f(\alpha) = 0
$.
It is the point at which 
$
\hat{\itPi}_\phi^{(n)}(\xi_n^*)
$
has the least $n$-dependence for large $n$.

\begin{figure}
  \includegraphics[height=.3\textheight]{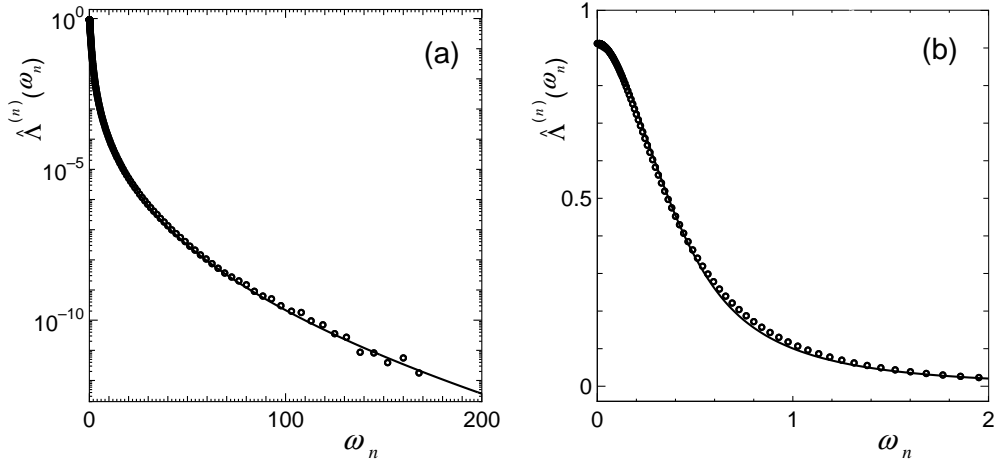}
  \caption{Comparison between the PDF of {\it fluid particle accelerations} 
measured in the DNS by Gotoh et al.\ at $R_\lambda = 380$ 
and the present theoretical PDF $\hat{\itLambda}^{(n)}(\omega_n)$ are
plotted on (a) log and (b) linear scales.
Closed circles are the DNS data points both on the left and right hand sides 
of the PDF. Solid lines represent the curves given 
by the present theory with $\mu = 0.240$ ($q=0.391$), 
$(n,\ \bar{n},\ q')=(17.5,\ 11.4,\ 1.70)$,
$
\omega_n^* = 0.622
$
($\alpha^* = 1.01$), and
$
\omega_n^{\rm max} = 2530
$.
\label{PDF acceleration log-linear Gotoh}}
\end{figure}

With the help of the second equality in (\ref{PDF kappa small}),
we obtain $\Delta \itPi_\phi^{(n)}(x_n)$, and have the formula 
to evaluate $\gamma_{\phi,m}^{(n)}$ in the form
$
2\gamma_{\phi,m}^{(n)} = \left(K_{\phi,m}^{(n)} - L_{\phi,m}^{(n)}\right) \Big/
\left(1 + K_{\phi,0}^{(n)} - L_{\phi,0}^{(n)}\right)
$
where
\bea
K_{\phi,m}^{(n)} \aeq \frac{3\ \delta_n^{\phi (m+1)\alpha^*/3 -\zeta_{2\phi}/2}}
{\phi \sqrt{2 \pi X \vert \ln \delta_n \vert}}
\int_0^1 dz\ z^{m} \left[1-\frac{1-q'}{2}\left(1+\frac{3f'(\alpha^*)}{\phi}\right)
\left( z^2 -1 \right) \right]^{1/(1-q')}
\\
L_{\phi,m}^{(n)} \aeq \frac{3\ \delta_n^{\phi m \alpha^*/3}}
{\phi \sqrt{2 \pi X \vert \ln \delta_n \vert}}
\int_{z_{\rm min}^*}^1 dz\ z^{m-1} 
\left[1 - \frac{1-q}{n}\ \frac{\left(3 \ln \vert z / z_0^* \vert \right)^2}{
2\phi^2 X \vert \ln \delta_n \vert} \right]^{n/(1-q)}
\eea
with
$
z_{\rm min}^* = \xi_{\rm min}/\xi_n^* 
=\delta_n^{\phi (\alpha_{\rm max} - \alpha^*)/3}
$,
$z_0^* = \xi_{n,0}/\xi_n^* 
=\delta_n^{\phi (\alpha_0 - \alpha^*)/3}
$.
We see that the tail part of the PDF, given by (\ref{PDF kappa large}), 
is mostly determined by 
the intermittency exponent $\mu$ and the multifractal depth $n$
which gives a length scale $\ell_n$.
On the other hand, the center part of the PDF, (\ref{PDF kappa small}),
is mainly controlled by $q'$.

The PDFs both for velocity fluctuations and for velocity derivatives are given by
the common formula $\hat{\Pi}^{(n)}(\xi_n) \equiv \hat{\Pi}_{\phi=1}^{(n)}(\xi_n)$ 
in their normalized variables $\xi_n = \delta u_n / \dbra (\delta u_n)^2 \dket^{1/2}$.
On the other hand, the PDFs both for pressure differences and for 
fluid particle accelerations are given by the common formula 
$\hat{\Lambda}^{(n)}(\omega_n) \equiv \hat{\Pi}_{\phi=2}^{(n)}(\omega_n)$ 
in their normalized variables $\omega_n = \delta p_n / \dbra (\delta p_n)^2 \dket^{1/2}$.
The PDF for energy dissipation rates is given with $\phi = 3$.

\begin{figure}
  \includegraphics[height=.25\textheight]{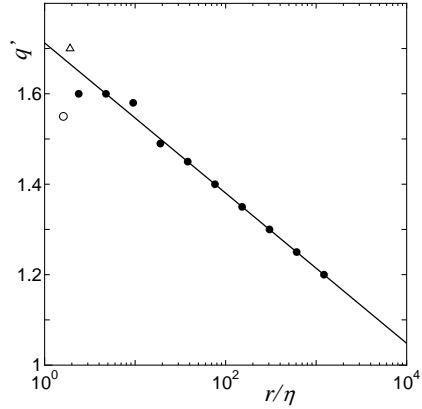}
  \caption{Dependence of $q'$ on the distance $r/\eta$ extracted from 
the analyses of the PDFs for velocity fluctuations (closed circles), 
for velocity derivatives (open circle) and for fluid particle accelerations 
(open triangle). The line represents $q'= -0.05 \log_2 (r/\eta) + 1.71$.
\label{qp-r}}
\end{figure}

The PDFs extracted by Gotoh et al.\ 
from their DNS data \cite{Gotoh02} at $R_\lambda = 380$ 
are shown, on log and linear scales, 
in Fig.~\ref{velocity fluctuations log} both for {\it velocity fluctuations} and 
for {\it velocity derivatives}, and in Fig.~\ref{PDF acceleration log-linear Gotoh} for 
{\it fluid particle accelerations}.
We found the value $\mu = 0.240$ by analyzing the measured scaling exponents 
$\zeta_m$ of velocity structure function with the formula (\ref{zeta}),
which gives the values $q = 0.391$, $\alpha_0 = 1.14$ and $X = 0.285$.
Through the analyses of the PDFs for velocity fluctuations in 
Fig.~\ref{velocity fluctuations log}, we extracted
the formula for the dependence of $n$ on $r/\eta$: \cite{AA6,AA7}
\bea
n \aeq -0.989 \times \log_2 r/\eta + 16.1
\quad (\mbox{for } \ell_c \leq r),
\label{n-roeta L larger} \\
n \aeq -2.40 \times \log_2 r/\eta + 24.0
\quad (\mbox{for } r < \ell_c).
\label{n-roeta L less}
\eea
This shows that the inertial range is divided into two scaling regions separated by 
the characteristic length $\ell_{\rm c} /\eta = 48.7$ which is close to 
the Taylor microscale $\lambda /\eta = 38.3$ of the system.
The equation (\ref{n-roeta L larger}) is consistent with the picture of 
the energy cascade model in which each eddy breaks up into 2 pieces at 
every cascade steps, whereas (\ref{n-roeta L less}) indicates that, for $r < \ell_c$,
each eddy breaks up, effectively, into $1.33 \approx 4/3$ \cite{AA7} pieces 
at every cascade steps.
This fact may be attributed to a manifestation of structural difference of eddies,
which can be checked by visualizing DNS eddies. Actually, one observes 
that DNS eddies with larger diameters than Taylor microscale $\lambda$ 
have rather round shapes, whereas eddies with smaller diameters 
compared with $\lambda$ have rather stretched shapes \cite{Tanahashi}.
The energy input scale for this DNS is estimated as the longest scale available
in the lattice with cyclic boundary condition, i.e.,
$\ell_{\rm in}/\eta = \pi/\eta \approx 1220$ with $\eta \approx 0.258 \times 10^{-2}$
\cite{Gotoh02} which gives the number of steps $\bar{n}$ 
within the energy cascade model through the formula (\ref{n-nbar}) with 
$\ell_0/\eta \approx 81300$ determined by (\ref{n-roeta L larger}).

For the analysis of the PDF for velocity derivatives 
in Fig.~\ref{velocity fluctuations log}, we chose 
the value $(n,\ \bar{n},\ q') = (22.4,\ 16.3,\ 1.55)$. 
The length corresponding to $n$ is calculated by (\ref{n-roeta L less})
to give  $r / \eta = 1.61$, 
which may provide us with an estimate for the effective shortest length in 
processing the DNS data to extract velocity derivatives.
Note that it is about the same order of the mesh size 
$\Delta r /\eta = 2\pi/(1024 \times \eta) \approx 2.38$ \cite{Gotoh02}
of the DNS lattice.

For the PDFs for {\it fluid particle accelerations} 
in Fig.~\ref{PDF acceleration log-linear Gotoh}, we have 
$(n,\ \bar{n},\ q') = (17.5,\ 11.4,\ 1.70)$.
Substitution of this value into (\ref{n-roeta L less}) gives 
the corresponding characteristic length $r/\eta = 7.91$ \cite{AA10}.
This may be the effective minimum resolution in cooking the DNS data to distill
accelerations.

In Fig.~\ref{qp-r}, we plotted the dependence of $q'$ on 
the distance $r/\eta$ extracted from the analyses of the DNS data, i.e.,
the closed circles are extracted from the PDF of velocity fluctuations,
the open circle is from the PDF of velocity derivatives and
the open triangle from the PDF of fluid particle accelerations.
The line represents 
$
q'= -0.05 \log_2 (r/\eta) + 1.71
\label{eq:qp-r}
$.
The points for $r/\eta > 20$ (closed circles) and 
for the accelerations (open triangle) are quite sensitive and easy to be determined.
Other points are insensitive and have a rather wide range in deciding the values $q'$.

\begin{figure}
  \includegraphics[height=.3\textheight]{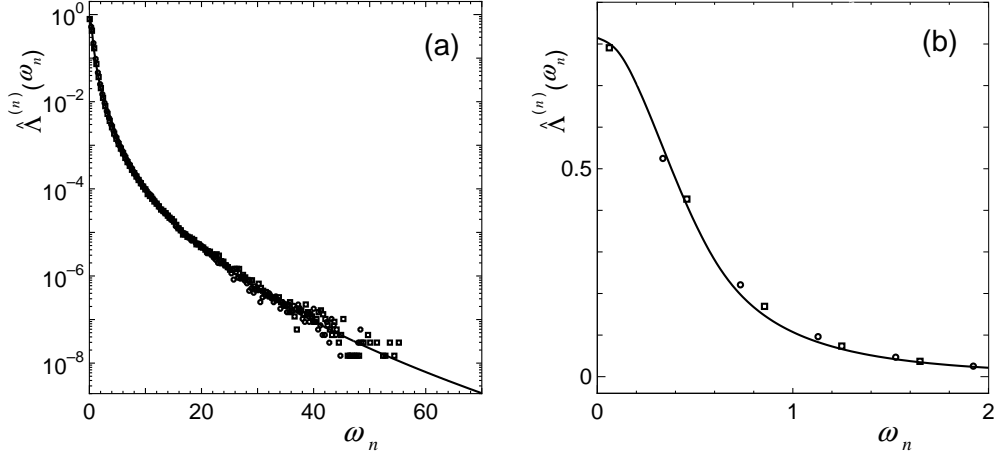}
  \caption{Comparison between the experimentally measured PDF of 
{\it fluid particle accelerations} by 
Bodenschatz et al.\ at $R_\lambda =690$ ($\rRe = 31\ 400$) 
and the present theoretical PDF $\hat{\itLambda}^{(n)}(\omega_n)$
are plotted on (a) log and (b) linear scales.
Open squares are the experimental data points on the left hand side 
of the PDF, whereas open circles are those on the right hand side. 
Solid lines represent the curves given 
by the present theory with $\mu = 0.240$ ($q=0.391$),
$(n,\ q')=(17.1,\ 1.45)$,
$
\omega_n^* = 0.605
$
($\alpha^* = 1.01$), and
$
\omega_n^{\rm max} = 2040
$.
\label{PDF acceleration log-linear Bodenschatz}}
\end{figure}

The PDFs for fluid particle accelerations 
measured by Bodenschatz et al.\ at $R_\lambda = 690$ \cite{EB02comment}
are given in Fig.~\ref{PDF acceleration log-linear Bodenschatz}.
We determined the value $n = 17.1$
by substituting the reported value $\ell_0=\ell_{\rm in}=$7.1~cm and 
the spatial resolution $0.5\ \mu\mbox{m}$ of the measurement for $\ell_n$
into its definition, 
$
n = \log_2 (\ell_0/\ell_n)
$.
The values $\mu = 0.240$ and $q' = 1.45$ are extracted by the analysis of
the experimental PDF with the derived theoretical formula \cite{AA10}.
Then, we have 
the values of parameters: $q = 0.391$, $\alpha_0 = 1.14$ and $X = 0.285$.
The flatness of the PDF turns out to be \cite{AA10}
$
F_{\mathrm{a}}^{(n)} \equiv \dbra \mathrm{a}_n^4 \dket
/\dbra \mathrm{a}_n^2 \dket^2
=\dbra \xi_n^4 \dket = 56.9
$
which is compatible with the value of the flatness $\sim 55\pm4$ 
reported in \cite{EB02comment}.

\section{Discussions and Prospects}

In this paper, the various experimental PDFs in turbulence
are analyzed precisely with the formulae 
(\ref{PDF kappa large}) and (\ref{PDF kappa small}) of the corresponding PDFs
derived by the MFA.
It is revealed that there are two distinct mechanisms underlying the dynamics of 
turbulence.
One contributes to the {\it tail} part of PDFs
and the other to the {\it center} part.
The structure of the tail part of the PDFs is determined by the global structure
representing the intermittent large deviations 
that is manifestations of the multifractal distribution of singularities 
in physical space due to the scale invariance of the Navier-Stokes equation 
for large Reynolds number. 
The specific form of the tail part comes from the assumption that
the probability to find a singularity exponent 
$\alpha$ within the range $\alpha \sim \alpha +d\alpha$ at a point in physical space 
is given by the Tsallis-type distribution function with the Tsallis parameter $q$.
The relation between $\alpha$ and an observing variable is given by 
the scale transformation.
On the other hand, the structure of the center part represents small deviations 
violating the scale invariance due to thermal fluctuations and/or observation error.
The center part is assumed to be given by the Tsallis-type distribution function 
with the Tsallis parameter $q'$ for the observing variable itself.
The value of $q'$ may be determined by a local structure of the system, e.g., 
the dynamics of a vortex, the mutual interaction between vortices and so on, and 
depends on the distance of two measuring points
in contrast to $q$. The latter parameter $q$ does not depend on the distance, 
and is determined once the value of the intermittency exponent is given.
It is one of the attractive future problems to derive two different dynamics 
taking care of the tail part controlled by $q$ and the center part by $q'$, 
and will be reported in the near future.

The success of the MFA in analyzing turbulence in high accuracy may provide us with
a good tool to see what is the origin of the singularities and 
why their distribution is multifractal. 
In order to investigate them, the vortex tangle is one of 
the attractive candidates as Feynman proposed \cite{Feynman}, 
since the vorticity in superfluid $^4$He and $^3$He is quantized and 
the normal component within the sense of the two fluid model can be negligible 
at very low temperature.
If the singularity originates from the core of vortex, the multifractality 
of turbulence in normal fluid can be related to various values 
of vorticities in the fluid. In this case, the vortex tangle may be uni-fractal,
and does not exhibit intermittency.
If the singularity originates from the reconnection of vortices, the multifractality 
of turbulence in normal fluid is related to the distribution 
of reconnection points in the fluid. Then, the vortex tangle may be also 
multifractal, and does exhibit intermittency.
A temperature-independent vortex decay mechanism below $T\sim70$ mK has been observed 
in superfluid $^4$He \cite{David00}, and the Kolmogorov spectrum (K41) is extracted 
from the simulation within the vortex filament model for non-frictional suferfluid 
$^4$He \cite{Tsubota02a}.
In tangle, the quantized vortex crosses the stream lines of superfluid velocity field, 
which may result in the decay mechanism at low temperature.
We expect that, through the analysis of the local dynamics controlled by 
the Tsallis parameter $q'$, we can extract some information 
about the mutual friction between the superfluid and normal components.

A circular vortex lattice is observed in the simulation of fast 
rotating Bose-Einstein condensate 
confined in a 2-dimensional quadratic-plus-quartic potential \cite{Tsubota02b}.
A possibility of generation of tangle phase in this system is one of the attractive 
problems.
The situation may be quite similar to the one in the generation of the Taylor-Couette 
turbulence in normal fluid.

Let us close this paper by mentioning that the vortex tangle can be 
an important stationary phase of {\it quLSI} (quantum LSI)
consisting of, for example, a huge number of superconducting loops of 
flux qubit \cite{flux qubit}.
The direction of current in a loop may change so quickly under 
an operation of quantum computer.
The congeries of the flux qubits can produce a tangle phase of 
flux quantums.
It may be important to see if the tangle phase benefits quantum entangled states 
or not.



\begin{theacknowledgments}
The authors would like to thank Prof.~R.H.~Kraichnan and 
Prof.~C.~Tsallis for their fruitful and enlightening comments 
with encouragement, and are grateful to Prof.~E.~Bodenschatz and Prof.~T.~Gotoh
for the kindness to show their data prior to publication.
\end{theacknowledgments}




\end{document}